\documentclass[review]{elsarticle}

\usepackage{lineno,hyperref}

\usepackage{epsfig,amsmath,hyperref}
\usepackage[english]{babel}
\usepackage{color}
\usepackage{amsfonts,amssymb,amsmath,amsopn,amsxtra}
\usepackage{mathrsfs}
\usepackage{enumerate}
\usepackage{bbm}
\usepackage{graphicx}
\usepackage{rotating}
\usepackage[normalem]{ulem}

\newcommand{\R}{\mathbb{R}}

\journal{Journal of \LaTeX\ Templates}

\begin{document}

\begin{frontmatter}

\title{Spectral and resonance properties of Smilansky Hamiltonian}

\author[mymainaddress,mysecondaryaddress]{Pavel Exner}
\ead[url]{gemma.ujf.cas.cz/~exner/}
\author[mymainaddress]{Vladimir Lotoreichik\corref{mycorrespondingauthor}}
\ead[url]{gemma.ujf.cas.cz/~lotoreichik/}
\author[mymainaddress]{Milo\v{s} Tater}
\ead{tater@ujf.cas.cz}

\cortext[mycorrespondingauthor]{Corresponding author}
\address[mymainaddress]{Department of Theoretical Physics, Nuclear Physics Institute, Czech Academy of Sciences, 25068 \v Re\v z near Prague, Czechia}
\address[mysecondaryaddress]{Doppler Institute for Mathematical Physics and Applied Mathematics, Czech Technical University, B\v rehov\'a 7, 11519 Prague,
Czechia}

\begin{abstract}
We analyze the Hamiltonian proposed by Smilansky to describe
irreversible dynamics in quantum graphs and studied further by
Solomyak and others. We derive a weak-coupling asymptotics of the
ground state and add new insights by finding the discrete spectrum
numerically. Furthermore, we show that the model has a rich
resonance structure.
\end{abstract}

\begin{keyword}
\texttt{Smilansky model \sep discrete spectrum \sep weak coupling
\sep resonances \MSC[2010] 35J10 \sep 35P15 \sep 65N25 \sep 81Q15
\sep 81Q35}
\end{keyword}

\end{frontmatter}


\section{Introduction}

While the fundamental equations of motion, both in classical and
quantum mechanics, are invariant with respect to time reversal, the
world around us is full of irreversible processes. On a microscopic
level, it is enough to recall spontaneous decays of particles,
nuclei, inelastic scattering processes, etc. And, of course, an
irreversible process \emph{par excellence} is the wave packet
reduction which is the core of Copenhagen description of a measuring
process performed on a quantum system.

A description of such processes in quantum mechanics is typically
associated with enlarging the state Hilbert space, conventionally
referred to as coupling the system to a heat bath \cite{Da, Ex}. As
a rule, it is assumed that (i) the bath is a system with infinite
number of degrees of freedom, (ii) the bath Hamiltonian has a
continuous spectrum, and (iii) the presence (or absence) of
irreversible modes is determined by the energies involved rather
than the coupling strength. While this all is true in many cases,
situations may exist where the system has neither of the listed
properties. This was the motivation which led Uzy Smilansky to
formulation a simple model \cite{Sm04} of a quantum graph coupled
to a heat bath consisting of a single harmonic oscillator which
exhibits an irreversible behavior. The easiest way to describe the
model is to phrase it in PDE terms, then its Hamiltonian is a
Schr\"odinger operator,
\begin{equation} \label{Hamiltonian}
H_\lambda = -\frac{\partial^2}{\partial x^2} + \frac12 \left(
-\frac{\partial^2}{\partial y^2} +y^2 \right) +\lambda y\delta(x)
\end{equation}
on $L^2(\R^2)$, with a singular interaction supported by the line
$x=0$ the strength of which depends on the coordinate $y$.

The system governed by the Hamiltonian \eqref{Hamiltonian} undergoes
an abrupt spectral transition at the critical value $\lambda =
\sqrt{2}$; while for $|\lambda| < \sqrt{2}$ the Hamiltonian is positive above this value its spectrum covers the whole real axis. The mechanism of this effect is easy to understand. The first and the last term in
\eqref{Hamiltonian} represent Hamiltonian of a one-dimensional
system with a point interaction which has a single eigenvalue equal
to $-\frac14 \lambda^2y^2\:$ \cite{AGHH}. This negative contribution
to the energy competes with the oscillator potential with the
balance flipping exactly at the said critical value of $\lambda$.

The model was subsequently an object of investigations,
generalizations and modifications in a series of mathematically
rigorous papers. The discrete spectrum in the subcritical case and
its behavior as $\lambda\to\sqrt{2}$ were analyzed in \cite{So04},
the absolute continuity of the other spectral component was
established in \cite{NS06}, an extension to the situation when the
`heat bath' consists of two or more oscillators was discussed in
\cite{ES05a, ES05b}. Other modifications consisted of replacing the
oscillator by a potential well of a different shape \cite{So06b} or
by replacing the line by a more general graph \cite{So06a}
corresponding to the original Smilansky proposal. It is also
possible to have the motion in the $x$ direction restricted to an
interval with periodic boundary conditions \cite{Gu11, RS07}; in the
first named of these papers time evolution of wave packets is
investigated confirming the idea that spectral change in the
supercritical case corresponds to the possibility of `escape to
infinity'. Let us also mention a modification in which the singular
interaction is replaced by a regular potential channel which gets
deeper and more narrow as the distance from a fixed point increases
\cite{BE14}.

In connection with the last mentioned system let us point out that
similar parameter dependent spectral transitions can be also
observed in other systems, an example being the one described by the
Hamiltonian
\begin{equation} \label{another}
H = -\Delta + |xy|^p - \lambda (x^2+y^2)^{p/(p+2)}\,, \qquad p\ge 1,
\end{equation}
on $L^2(\R^2)\;$ \cite{BEKT15}. In this case the change is even more
dramatic, from the spectrum which is below bounded and purely
discrete to the whole real line, the transition occurring at the
critical value $\lambda=\lambda_p$, the ground state eigenvalue of
the appropriate (an)harmonic oscillator. To our knowledge, spectral
transitions of this type were for the first time noted in
\cite{Zn98}.

Our object of interest in the present paper is the original
Smilansky Hamiltonian \eqref{Hamiltonian}. Despite the fact that
many of its properties were demonstrated in the papers mentioned
above, there is still a room for new insights. The first topic here
concerns the discrete spectrum of $H_\lambda$ and comes from the
observation that it is relatively easy to find the eigenvalues and
eigenfunctions numerically turning the task into a matrix problem of
a sufficiently large dimension.

Another question to be discussed here concerns \emph{resonances} in
the system described by the Hamiltonian \eqref{Hamiltonian}. While
the absolute continuity of the spectrum above the threshold and the
existence of the wave operators were established in \cite{ES05a},
the possible resonance character of the scattering seems to have
escaped attention. In our recent paper \cite{ELT16} we have
demonstrated how one can find poles of the analytically continued
resolvent and derived asymptotic expansions for them, here we focus
on the physical meaning of the resonances and illustrate their
behavior by numerical results.

\section{Survey of results about the spectrum}

To make the paper self-contained we summarize first the known facts
about the operator that are indicated briefly in the introduction. It is
obvious that we can consider the positive values of the coupling constant
$\lambda$ as the opposite case is obtained via a mirror
transformation with respect to the axis $x=0$.

\emph{(a) Spectral transition:} In the subcritical case,
$\lambda<\sqrt{2}$, the particle remains localized in the vicinity of
the $x$ axis, while for $\lambda>\sqrt{2}$ it can according to
\cite{Gu11} escape to infinity along the singular `channel' in the
$y$ direction. In spectral terms, this corresponds to the switch
from a positive spectrum of $H_\lambda$ to a below unbounded one
which occurs at $\lambda=\sqrt{2}$.

\emph{(b) The continuous spectrum} of $H_\lambda$ covers the
interval $(\frac12,\infty)$ for $\lambda<\sqrt{2}$,
the positive semi-axis for $\lambda = \sqrt{2}$, and the whole
real axis if $\lambda>\sqrt{2}$. Moreover, this spectrum is in fact
purely absolutely continuous \cite[Sec.~VII.2]{RS1}, in particular, the
operator is purely absolutely continuous in the supercritical case
when $\sigma(H_\lambda) = \sigma_\mathrm{ac}(H_\lambda) = \R$.

\emph{(c) Eigenvalue existence and location:} For any $\lambda\in(0,\sqrt{2})$ the discrete spectrum of $H_\lambda$ is nonempty, simple, and finite, being contained in $(0,\frac12)$, i.e. the subcritical operator is positive. The eigenvalues are decreasing functions of the coupling constant $\lambda$. On the other hand, there are no eigenvalues $\ge\frac12$, and in the supercritical case, $\lambda>\sqrt{2}$, the point spectrum of $H_\lambda$ is empty, in other words, there are no eigenvalues embedded in the continuum.

\emph{(d) Near critical asymptotics:} The number of eigenvalues in the subcritical regime increases with the coupling constant and explodes as it approaches the critical value $\lambda=\sqrt{2}$ behaving asymptotically as \cite{So04}
 \begin{equation} \label{solombound}
	N(\textstyle{\frac12},H_\lambda) \sim \frac14 \sqrt{\frac{1}{\sqrt{2}(\sqrt{2}-\lambda)}}
 \end{equation}
where $N(\mu,A)$ conventionally denotes the number of eigenvalues of the operator $A$ below $\mu$ with the multiplicity taken into account.

 \emph{(e) Weak-coupling asymptotics:} For small enough $\lambda$ there is a single eigenvalue which behaves as
 \begin{equation} \label{weak}
\epsilon_1(\lambda) = \frac12 - \frac{\lambda^4}{64} + \mathcal{O}(\lambda^5)
 \end{equation}
as $\lambda\to 0\;$ \cite{ELT16}. In fact, due to the mirror symmetry of the spectrum with respect to $\lambda$ the error term can be replaced by $\mathcal{O}(\lambda^6)$.

\section{Numerical solution of the eigenvalue problem}

To find the discrete spectrum we note first that the solutions to the eigenvalue problem can be expanded using the `transverse' base naturally spanned by the functions
\begin{equation} \label{ho}
\psi_n(y) = \frac{1}{\sqrt{2^n n! \sqrt\pi}}\: \mathrm{e}^{-y^2/2} H_n(y)
\end{equation}
corresponding to the oscillator eigenvalues $\nu_n=n+\frac12,\: n=0,1,2,\dots\,$; here $H_n$ are Hermite polynomials with conventional
normalization.
Furthermore, the task can be simplified by noting that the problem has a mirror symmetry w.r.t. \mbox{$x=0$} which allows us to decompose $H_\lambda$ into the trivial odd part $H_\lambda^{(-)}$ and the even part $H_\lambda^{(+)}$ which is equivalent to the operator on $L^2(\R^2_+)$, where $\R^2_+:= \mathbb{R}\times(0,\infty)$, with the symbol \eqref{Hamiltonian} and the domain consisting of $H^2(\R^2_+)$ functions satisfying the boundary conditions
\begin{equation} \label{bc}
	f_x^\prime(0+,y) = \frac12\, \lambda y f(0+,y)\,.
\end{equation}
Substituting the Ansatz
\begin{equation} \label{Ansatz}
f(x,y) = \sum_{n=0}^\infty c_n\, \mathrm{e}^{-\kappa_n x} \psi_n(y)
\end{equation}
with
\begin{equation} \label{kappa}
\kappa_n:= \sqrt{n+\frac12-\epsilon}
\end{equation}
into \eqref{bc} and using orthonormality of the basis \eqref{ho}, we get for solution with the energy $\epsilon$ the equation
\begin{equation} \label{secular}
B_\lambda c = 0\,,
\end{equation}
where $c$ is the coefficient vector and $B_\lambda$ is an operator in $\ell^2$, in other words, an infinite matrix with the elements
\begin{equation} \label{sec_elem}
(B_\lambda)_{m,n} = \kappa_n \delta_{m,n} + \frac12\lambda (\psi_m,y \psi_n)\,;
\end{equation}
note that the matrix $B_\lambda$ is in fact tridiagonal because
$$
(\psi_m,y \psi_n) = \frac{1}{\sqrt{2}} \big( \sqrt{n+1}\, \delta_{m,n+1} + \sqrt{n}\, \delta_{m,n-1} \big)\,,
$$
in other words, $B_\lambda$ is a Jacobi matrix.

Note that using a general sequence $\{u_n(x)\}$
instead of $\mathrm{e}^{-\kappa_n x}$ in the Ansatz~\eqref{Ansatz} one
can reformulate the problem as an analogue of the Birman-Schwinger principle as first proposed by M.~Solomyak in \cite{So04}, see a detailed discussion in~\cite{ELT16}. The same approach works also for some modifications of the original Smilansky model mentioned in the introduction where, however, the structure of the appropriate matrix problem is different and may be more complicated.

To find the values of $\epsilon$ for which the secular equation \eqref{secular} is satisfied for a given $\lambda$ we solve a truncated matrix problem increasing the cut-off until a numerical stability is reached. The convergence is slower the closer is the value of $\epsilon$ to the threshold value $\frac12$. We present results for the sizes up to $9000\times9000$.

\begin{figure}
\hspace{0em}\includegraphics[angle=0,width=12cm]{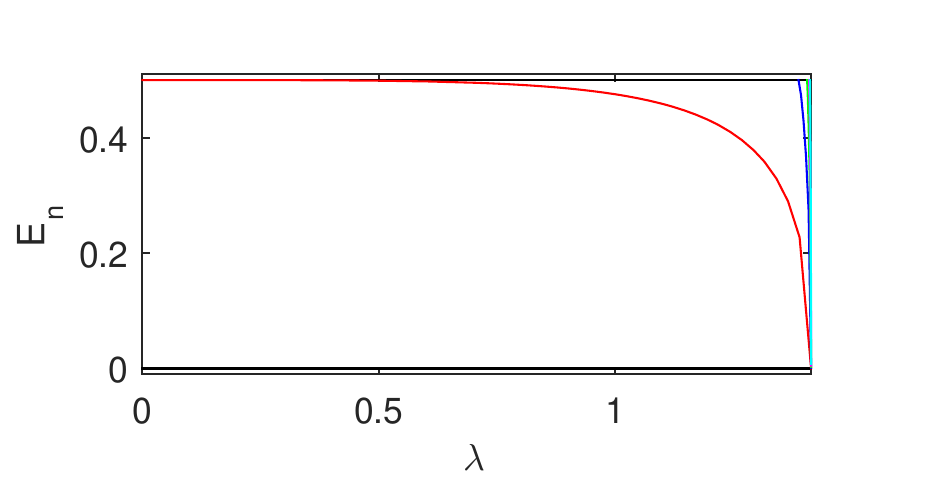}
\caption{Eigenvalues of $H_\lambda$ as functions of $\lambda$}
\label{ev_plot}
\end{figure}

The results are plotted below in a series of figures. They are, of course, in agreement with the theoretical predictions listed in the previous section, but they provide additional insights. To begin with, Fig.~\ref{ev_plot} shows eigenvalues of $H_\lambda$ as functions of the coupling constant. We see that in the most part of the subcritical region, more than $98\%$ of it, the Hamiltonian has a single eigenstate. The second eigenvalue appears only at $\lambda\approx 1.387559$, the next thresholds are $1.405798,1.410138,1.41181626,1.41263669, \dots$.
\begin{figure}
\includegraphics[angle=0,width=10cm]{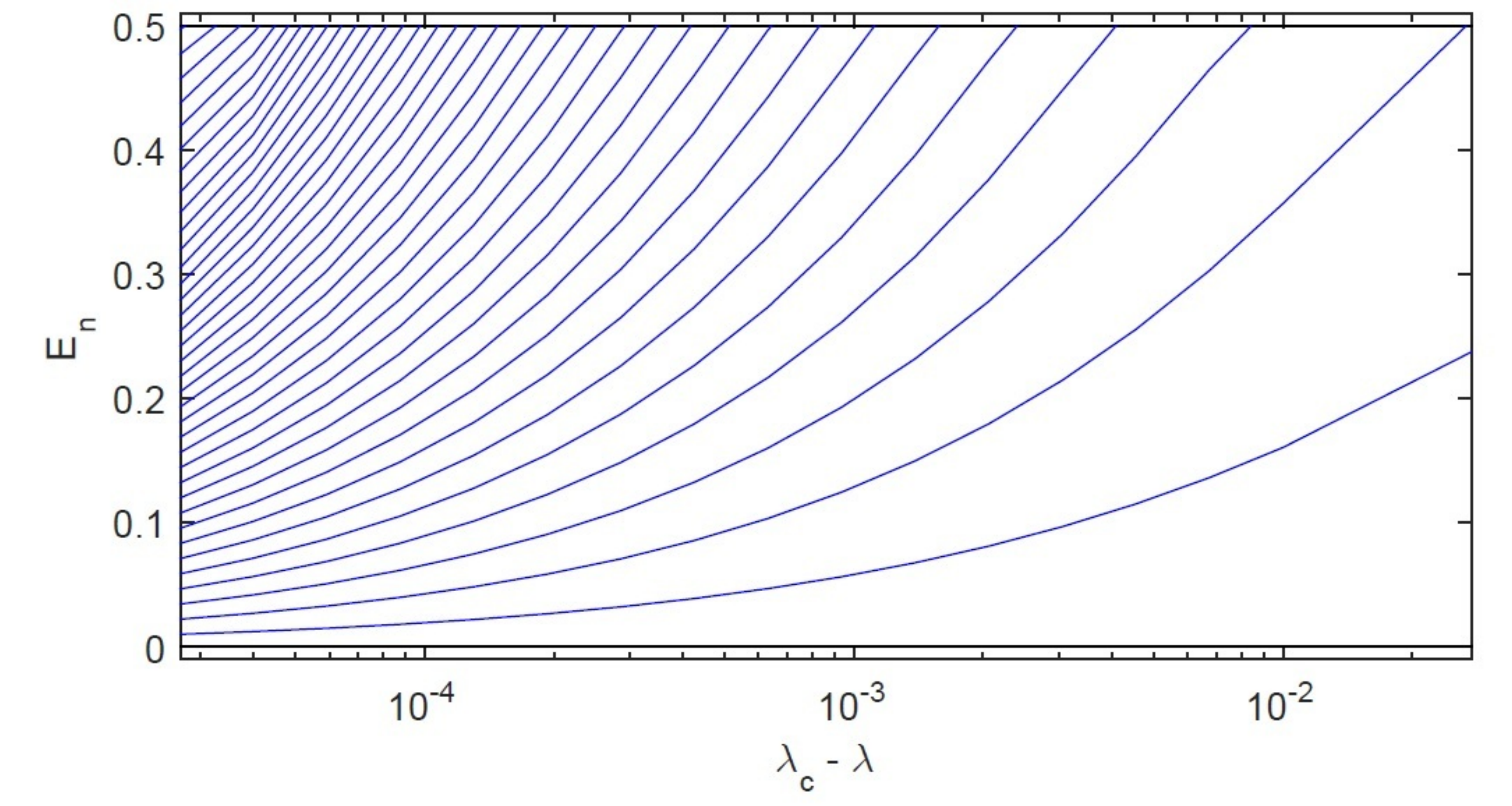}
\caption{The discrete spectrum near the critical point}
\label{log_plot}
\end{figure}
This does not give a good picture of the spectral behavior near the critical values. Instead we can plot the eigenvalue as functions of $\lambda$ using a logarithmic scale. The result is shown in Fig.~\ref{log_plot}. We see that their number indeed grows as we approach the transition point filling the allowed interval so that $\sigma(H_{\sqrt{2}})=[0,\infty)$, but that the discrete spectrum is asymptotically equidistantly distributed.
\begin{figure}
\includegraphics[angle=0,width=10cm]{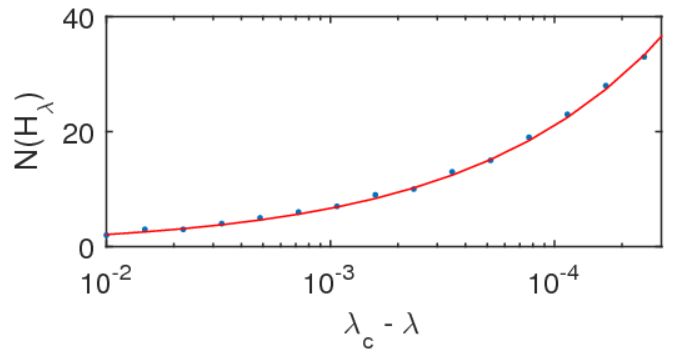}
\caption{The eigenvalue numbers (indicated by dots) compared with the asymptotics \eqref{solombound}}
\label{solomfig}
\end{figure}
The same result allows us to illustrate the Solomyak asymptotic formula \eqref{solombound}: it is obvious from Fig.~\ref{solombound} that the leading term of the expansion gives a good approximation as long as $\sqrt{2}-\lambda \lesssim 10^{-2}$.

\begin{figure}
\includegraphics[angle=0,width=10cm]{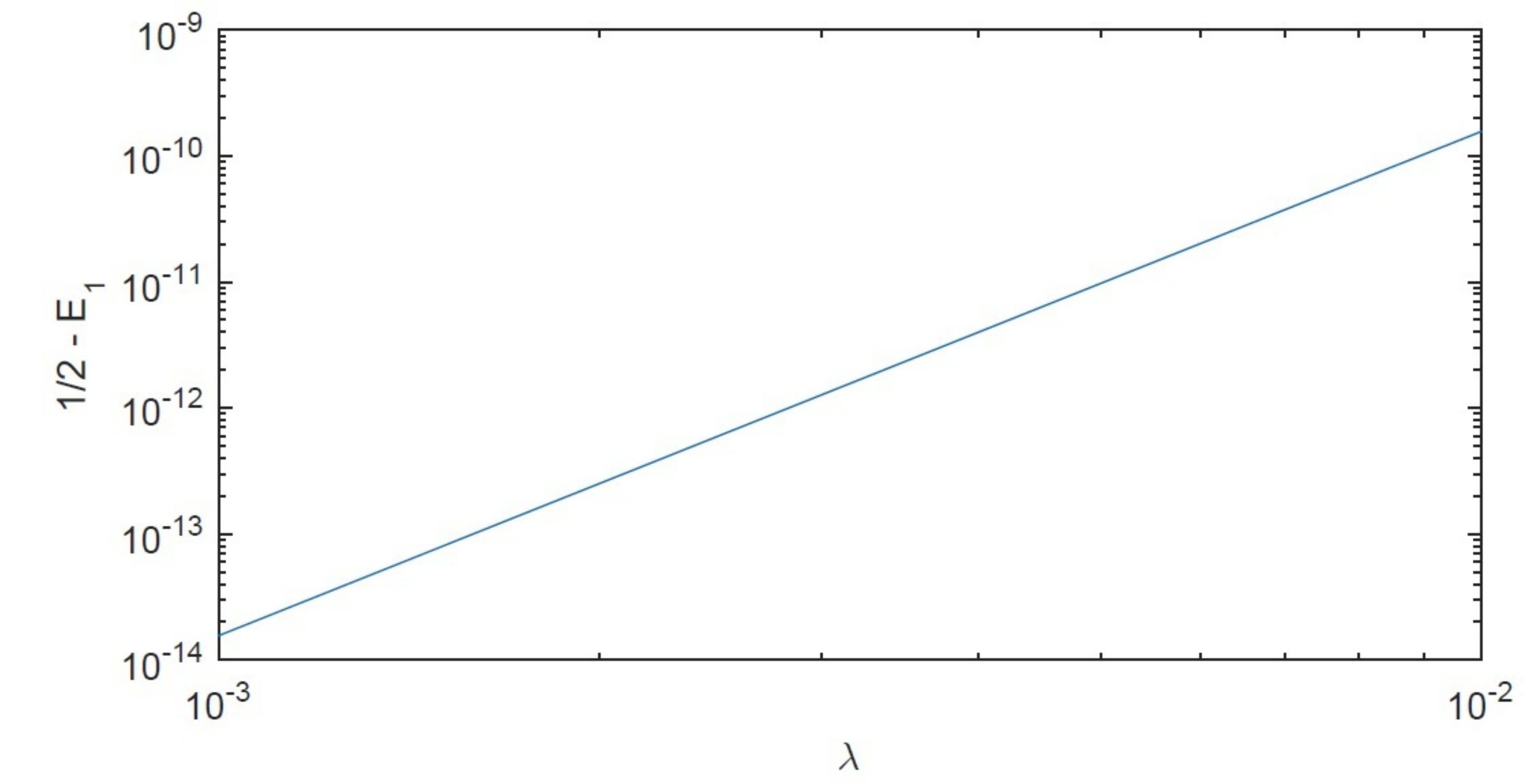}
\caption{The weak-coupling behavior of the ground-state energy}
\label{weakfig}
\end{figure}

The opposite asymptotical situation is the weak coupling. In Fig.~\ref{weakfig} we plot the ground-state eigenvalue as a function of $\lambda$. In the double logarithmic scale the graph is practically indistinguishable from a straight line which allows us to read out the power of the leading term as well as the coefficient $0.015625 = \frac{1}{64}$ appearing in the leading term of \eqref{weak}.

Since, in contrast to points (a)--(d) of the previous section the weak-coupling asymptotics is a new observation it is worth to show a way how it can be derived. To this purpose we introduce a symbol for the energy gap
\begin{equation} \label{weakgap}
    \mu_\lambda := 1/2 - \epsilon_1(\lambda)
\end{equation}
and denote by $c^\lambda = (c^\lambda_0, c^\lambda_1, c^\lambda_2,  \dots)$ the normalized eigenfunction of $B_\lambda$ given by \eqref{secular} corresponding to zero eigenvalue. Without loss of generality we may assume that
$c^\lambda_0$ is non-negative.
Writing the lines of the corresponding infinite system of equation explicitly, we get
\begin{equation}\label{weakJacobi}
\begin{split}
    & \sqrt{\mu_\lambda}\,c^\lambda_0 + \frac{\lambda}{2\sqrt{2}}\,c^\lambda_1 = 0\,, \\[.5em]
    & \frac{\sqrt{k}\lambda}{2\sqrt{2}}\,c^\lambda_{k-1} + \sqrt{k + \mu_\lambda}\,c^\lambda_k
       + \frac{\sqrt{k+1}\lambda}{2\sqrt{2}}\,c^\lambda_{k+1} = 0\,, \qquad k\ge 1\,.
\end{split}
\end{equation}
Using elementary inequalities $\sqrt{k}\le \sqrt{k+\mu_\lambda}$ and that $\sqrt{k+1}\le\sqrt{2(k+\mu_\lambda)}$, we get from the second relation in \eqref{weakJacobi} the estimate
\begin{equation}\label{ineq}
    |c^\lambda_k| \le
    \frac{\lambda}{2\sqrt{2}}\, |c^\lambda_{k-1}| + \frac{\lambda}{2}\, |c^\lambda_{k+1}|\,,
    \qquad k\ge 1\,,
\end{equation}
which implies
$$
    |c^\lambda_k|^2 \le
    \frac{\lambda^2}{4}\, |c^\lambda_{k-1}|^2 + \frac{\lambda^2}{2}\, |c^\lambda_{k+1}|^2\,,
    \qquad k\ge 1.
$$
and combining it with the normalization assumption, $\| c^\lambda \| = 1$, we arrive at
$$
    \sum_{k=1}^\infty |c^\lambda_k|^2
    \le
    \frac{\lambda^2}{4} \sum_{k=0}^\infty\, |c^\lambda_k|^2 +
    \frac{\lambda^2}{2} \sum_{k=2}^\infty\, |c^\lambda_k|^2 \le \frac{3\lambda^2}{4}.
$$
Using $\| c^\lambda \| = 1$ once more, we conclude that
\begin{equation}\label{bst0}
    c^\lambda_0 = \Big(\|c^\lambda\|^2 - \sum_{k=1}^\infty\, |c^\lambda_k|^2\Big)^{1/2}
       = 1 + O(\lambda^2)\,, \qquad \lambda \rightarrow 0\,;
\end{equation}
then inequality \eqref{ineq} together with the normalization yield $c^\lambda_2 = \mathcal{O}(\lambda)$. In combination with~\eqref{bst0} and the second relation of~\eqref{weakJacobi} for $k = 1$ this implies further
\begin{equation}\label{bst1}
    c^\lambda_1 = \frac{\lambda}{2\sqrt{2}} + \mathcal{O}(\lambda^2)\,,\qquad \lambda\rightarrow 0\,.
\end{equation}
Using now~\eqref{bst0} and~\eqref{bst1} we obtain from the first relation in~\eqref{weakJacobi} that
$$
    \mu_\lambda = \frac{\lambda^4}{64} + \mathcal{O}(\lambda^5)\,,
    \qquad
    \lambda\rightarrow 0\,,
$$
which is nothing else but the formula \eqref{weak}. In the counterpart to this paper \cite{ELT16} we give a modified version of this argument which allows, at least in principle, to compute higher terms of this asymptotic expansion.

\begin{figure}
\vspace{-6em}
\includegraphics[angle=0,width=12cm]{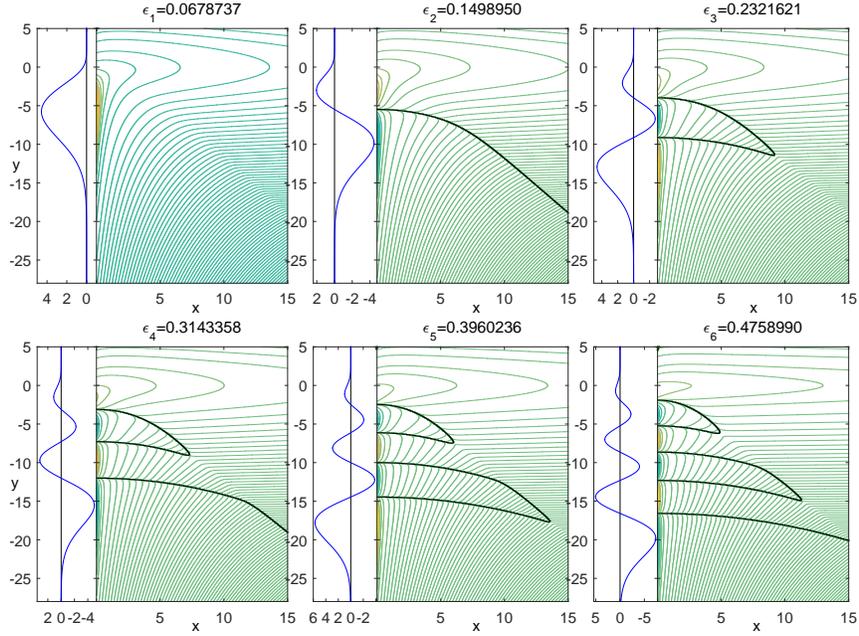}
\vspace{-8em}
\caption{The first six eigenfunctions of $H_\lambda$ for $\lambda=1.4128241$}
\label{eigenfig}
\end{figure}
Before proceeding further, let us also look at the eigenfunctions. In Fig.~\ref{eigenfig} we show a contour plot of the first six eigenfunctions of $H_\lambda$ for $\lambda=1.4128241$, in other words, $\lambda=\sqrt{2}-0.0086105$. 
In the side plots we feature restrictions of the eigenfunctions
on the $y$-axis. As expected the eigenfunctions are concentrated in the vicinity of the `escape channel', i.e. the negative part of the $y$-axis. The plots also show that they obey the usual restriction, commonly referred to as Courant nodal domain theorem; the inequality becomes sharp from the third eigenfunction. The pattern is interesting, the odd eigenfunctions
(except the ground state) have closed nodal lines only, the even ones
have an unbounded nodal line. One wonders whether this property holds for the discrete spectrum generally.

\section{Resonances}

Our second main point is to show that Smilansky model also exhibits interesting resonance behavior. Having said that, one has to explain first what is meant by a resonance. There are different definitions, the two main being the \emph{resolvent resonances} associated with poles in the analytic continuation of the resolvent over the cut(s) corresponding to the continuous spectrum, and \emph{scattering resonances} associated with singularities of the scattering matrix. While these two can often be identified, this property has to be checked in each particular case; recall that the former are determined by a single operator, the latter come from comparison of the free and interacting dynamics.

From the differential equation point of view finding poles of the analytically continued resolvent means to solve the eigenvalue equation for looking for a complex `eigenvalue' corresponding to an `eigenfunction' which does not belong to $L^2$ being exponentially increasing with $|x|$. Modifying the argument of the previous section which led to the infinite matrix problem the task is again reduced to the equation \eqref{secular}. A subtle point in this case, as with other `guiding channel' systems with infinite number of transverse modes \cite{EK} producing infinite number of cuts in the continuous spectrum, is that the Riemann surface of energy has infinite number of sheets, in fact an uncountable one. We are interested in those of them neighboring with the physical sheet which in the present context means that looking for resonances on the $n$-th sheet we have to flip sign of the first $n-1$ signs of the square roots in \eqref{secular}; for a more thorough discussion of this problem we refer the reader to~\cite{ELT16}.

Let us turn to scattering resonances. Suppose that the incident wave comes in the $m$-th channel from the left with the energy $k^2$. We use the Ansatz
\begin{equation} \label{scattAnsatz}
	f(x,y) =
	\begin{cases} \sum_{n=0}^\infty \Big( \delta_{mn}
	\mathrm{e}^{-ip_m x} \psi_n(y) + r_{mn}\, \mathrm{e}^{ip_nx} \psi_n(y)\Big),
	&\quad x < 0, \\[.5em] \sum_{n=0}^\infty t_{mn}\, \mathrm{e}^{-ip_nx} \psi_n(y),
	&\quad x > 0. \end{cases}
\end{equation}
where we have denoted $p_n = p_n(k) := \sqrt{k^2 -\nu_n}$. Let us note that the on-shell scattering matrix with the elements has the size $M\times M$ where $M:=\big[k^2-\frac12\big]$, the other `reflection' and `transmission' amplitudes correspond to the evanescent modes present in the scattering solution. The sign choice of the square roots signifying the Riemann sheet choice is the same as in the previous case.

It is straightforward to compute from here the boundary values $f(0\pm,y)$ and $f'_x(0\pm,y)$. The continuity requirement at $x=0$ together with the orthonormality of the basis $\{\psi_n\}$ yields
\begin{equation} \label{rt}
t_{mn} = \delta_{mn} + r_{mn}\,.
	\end{equation}
Furthermore, substituting the boundary values from the Ansatz \eqref{scattAnsatz} into
\begin{equation} \label{jump}
f'_x(0+,y)-f'_x(0-,y)-\lambda yf(0,y) = 0
\end{equation}
and integrating the obtained expression with $\int \mathrm{d}y\, \psi_l(y)$ we obtain
\begin{equation} \label{scttsec}
\sum_{n=0}^\infty \Big( 2p_n \delta_{ln} -i\lambda(\psi_l,y\psi_n) \Big) r_{mn} = i\lambda(\psi_l,y\psi_m)\,.
\end{equation}
The requirement of solvability of this system leads to the same condition as before which allows us to conclude that the \emph{resolvent and scattering resonances in Smilansky model coincide}.

The next question concerns the existence of resonances and their behavior as the coupling constant $\lambda$ changes. There are some standard mechanisms producing resonances. One is a perturbation of embedded eigenvalues of the Hamiltonian, however, we have seen that in the present case there are no such eigenvalues. Another possibility comes from perturbation of the singularities associated with the thresholds of the transverse channels, i.e. the branching points of the energy surface cuts

The weak-coupling analysis follows the route as for the discrete spectrum and shows that for small $\lambda$ a resonance pole splits of each threshold according to the asymptotic expansion. This indeed leads to emergence of complex poles in the vicinity of thresholds which behave as
\begin{equation} \label{weakres}
\rho_m(\lambda) = - \frac{\lambda^4}{64} \big( 2m+1 + 2im(m+1)\big) + \mathcal{O}(\lambda^5)
\end{equation}
as one can readily check repeating the argument of the previous section, in fact, the weak coupling of the bound state is included if we put $m=0$ in \eqref{weakres}. For $m=1,2,\dots\,$ we get genuine complex poles the distance for the corresponding threshold is proportional to $\lambda^4$ and the trajectory asymptote is the `steeper' the larger $m$ is. A more detailed discussion of the weak-coupling behavior of these resonances
and also of resonances lying on the sheets that are not adjacent
to the physical sheet can be again found in \cite{ELT16}.

\begin{figure}
\vspace{0em}
\begin{center}
\includegraphics[angle=0,width=10cm]{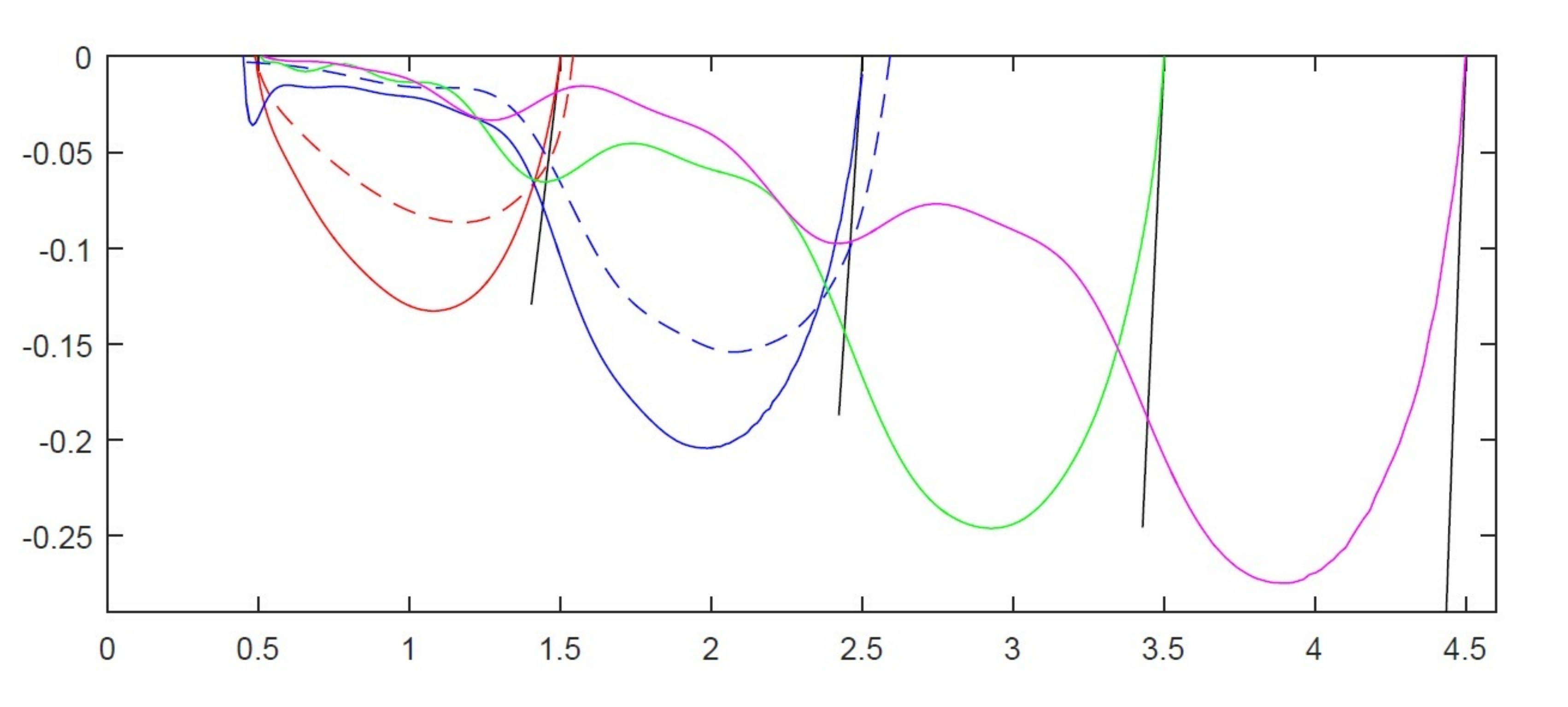}
\caption{Resonance trajectories as $\lambda$ runs through the
interval $(0,\sqrt{2})$. Resonances that do not depart from
thresholds are marked by dashed lines} \label{resofig}
\end{center}
\end{figure}

However, this tells us nothing about the pole behavior for larger
$\lambda$. To get a more complete picture, we again analyze the
problem numerically.

As described above, we look for complex values of $\epsilon$ for which the secular equation \eqref{secular} is satisfied for a given $\lambda$. 
Now, however, $B_\lambda$ is modified by flipping sign of the first $n-1$ signs of the square roots. Then, a truncated matrix problem ia solved,
tuning the cut-off until a numerical stability is reached.

The results are shown in Fig.~\ref{resofig} where we demonstrate resonances
splitting from the lowest thresholds, from the second to the fifth
one together with their weak-coupling asymptotes according to
\eqref{weakres}. We see that the behavior beyond the weak-coupling
regime is different, the pole trajectories return to the real axis
and they may even end up as an isolated eigenvalue. This is not
surprising, however, since a similar behavior is known from other
resonance models \cite{Ex13} being first observed in the classical
Lee-Friedrichs model \cite{Ho58}.

What is even more interesting, the numerical analysis shows that the
system has other resonances that do not emerge from thresholds and
appear when the coupling surpasses a critical value. In
Fig.~\ref{resofig} we show two such trajectories located at the
second and third sheet of the energy surfaces and emerging at
$\lambda=1.287$ and $\lambda=1.19$, respectively. One can conjecture
naturally that these are not the only ones and that the system
governed by Smilansky Hamiltonian has a rich family of resonances of
different kinds which deserve a separate investigation.

\section*{Acknowledgments}

\noindent This research was supported by the Czech Science Foundation (GA\v{C}R) within the project 14-06818S.


\end{document}